**Distinguishing Step Relaxation Mechanisms via Pair Correlation Functions**

D.B. Dougherty, I. Lyubinetsky[†], T.L. Einstein, E.D. Williams*

Department of Physics and MRSEC

University of Maryland

College Park, MD 20742-4111

Theoretical predictions of coupled step motion are tested by direct STM measurement of the fluctuations of near-neighbor pairs of steps on Si(111)-$\sqrt{3} \times \sqrt{3}$ R30º Al at 970K. The average magnitude of the pair-correlation function is within one standard deviation of zero, consistent with uncorrelated near-neighbor step fluctuations. The time dependence of the pair-correlation function shows no statistically significant agreement with the predicted $t^{1/2}$ growth of pair correlations via rate-limiting atomic diffusion between adjacent steps. The physical considerations governing uncorrelated step fluctuations occurring via random attachment/detachment events at the step edge are discussed.

[†] Permanent Address: Pacific Northwest National Lab, P.O. Box 999, K8-93, Richland WA 99352

* Corresponding Author. Tel.: +1-301-405-6156; fax:+1-301-314-9465; e-mail : edw@physics.umd.edu.

Structural and compositional fluctuations will play an increasingly important role in thermal stability, device performance, and fabrication processes as material structures shrink to the nanoscale, and surface-to-volume ratios increase [1-3].  On solid crystalline surfaces and interfaces, crystalline-layer boundaries (steps) undergo thermal fluctuations [4,5], which can couple into charge carrier scattering [6] and structural evolution [7].  The ability to observe step fluctuations in real time using scanned probe and electron microscopies provides the opportunity to quantify this stochastic behavior and to understand the underlying mechanisms.

For more than a decade it has been recognized that the statistical properties of step fluctuations are determined by the rate-limiting transport mechanism underlying their relaxation [8,9].  When the stochastic time evolution of step edge position x(t) can be accessed, it is possible to determine the temporal correlation function [4,5]

$$G(t) = \langle [x(y_0, t+t_0) - x(y_0, t_0)]^2 \rangle, \qquad (1)$$

which quantifies the mean-squared displacement (i.e. fluctuation) of the step as a function of time.  When the underlying microscopic processes governing the relaxation have distinct energies, the temporal correlation function is expected to grow with a fractional power of time,

$$G(t) = ct^{\frac{1}{z}} \qquad (2)$$

with an exponent, $z$, that depends only on the rate-determining transport mechanism and a prefactor that depends on the energetic and kinetic parameters of the underlying microscopic processes.

Power-law behavior has been observed experimentally for steps on metal [5,10] and semiconductor surfaces [4], in virtually all cases yielding exponents of $z=2$ or $z=4$. For an isolated step, analytical results for $G(t)$ exist for three important limiting cases of surface transport processes [1,2]. These are 2D evaporation-condensation (EC) with $z=2$, step-edge diffusion (SED) with $z=4$, and terrace diffusion (TD) with $z=3$. In addition, if one considers steps that are not isolated, e.g. steps on a vicinal surface with an average step-step spacing $L$ [11], the power-laws of the correlation functions are predicted not to change for evaporation-condensation and step edge diffusion[12-14]. However, when terrace diffusion is rate-limiting, two different regimes of correlation-function scaling can occur. For step fluctuations that are much smaller in wavelength (and thus amplitude) than the spacing between steps, $z=3$ behavior is still predicted. For longer wavelength fluctuations, a crossover to $z=2$ scaling occurs. This diffusion-from-step-to-step (DSS) mechanism [8] thus yields an alternative explanation for the observation of an exponent $z=2$. Fig. 1 shows schematically the two surface processes leading to $z=2$ overlaid on a 3D rendering of a real pair of monatomic steps. In both 2D-EC and DSS, step relaxation occurs by the exchange of mass with a reservoir. For EC, the reservoir is provided by the equilibrium concentration of the mobile species on the terraces between the steps. In this case, terrace diffusion must occur so rapidly that any fluctuations in concentration due to step motion are healed by diffusion much faster than the time scale of attachment and detachment events at the steps themselves. In contrast, the reservoir for the DSS case is provided by the neighboring steps, which act as perfect sinks and sources of mobile adspecies. Microscopically, this involves rate-limiting transit over the terraces, and a relative attachment /detachment rate so fast that it is essentially instantaneous. Before moving on, we must note that mass transport in any real system rarely involves just one of the ideal processes, and instead always represents a balance of all three [12,13]. In this more complete picture, the wavelength of fluctuations serves as an important parameter in determining which underlying mechanism dominates the measured correlation function.

Scaling of the correlation function with $z=2$ (or equivalently, scaling of Fourier mode decay time with the square of wave number [8-10, 12-14,16-17]) has been observed in experiment for Si(111) [8,15] at elevated temperatures, for Si(001) in two distinct temperature regimes [16,17], for the corrugated metal surfaces Au(110) [18] and Mo(011) [10] at elevated temperature and Pd (110) [19] at room temperature, and for Si(111) with a $\sqrt{3} \times \sqrt{3}$ R30° overlayer of Al at elevated temperature [20]. It is rather surprising that the $z=3$ signature of terrace diffusion for isolated steps (or short wavelength fluctuations) has scarcely ever been seen. It is possible that in some cases this is due to a DSS mechanism, which requires long-wavelength fluctuations, and in others to an attachment/detachment process (2D-EC), which can be present independent of wavelength. In this work we demonstrate the application of a second method of analysis of step fluctuations, the step-step pair correlation function, to distinguish the two physical mechanisms that can yield $z=2$ in the temporal correlation function.

Consider first the temporal correlation function $G(t)$: if the steps in an array are far apart relative to the length-scale of step fluctuations, then the isolated step interpretation of a rate-limiting 2D-EC for $z=2$ is valid even when 2D-evaporation/condensation and terrace diffusion occur simultaneously. The form of $G(t)$ is then given by [4,5,12,13],

$$G_{EC}(t) = \sqrt{\frac{4k_B T a_n^2 a_p t}{\pi \tilde{\beta} \tau_a}} \qquad (3)$$

where $\tau_a$ is the average time between EC events at the step edge, $\tilde{\beta}$ is the step stiffness, and $a_n$ and $a_p$ are the projected surface lattice parameters normal and parallel to the step, respectively. If the steps are closely spaced relative to the wavelength of step fluctuations, diffusion across the terraces between two neighboring steps, with fast evaporation/condensation processes (relative to the terrace diffusion rate) at the step edges themselves results in similar power-law scaling with $z=2$ [2,3,6-9]. In this DSS case, $G(t)$ becomes [6,7],

$$G_{DSS}(t) = \sqrt{\frac{16 D c k_B T \Omega^2 t}{\pi^3 \tilde{\beta} L}}, \qquad (4)$$

where $L$ is the step separation and the kinetic parameter appearing in the expression is the combination of the surface diffusion coefficient $D$ and the equilibrium surface adspecies concentration $c$.

To distinguish 2D-EC from DSS when the correlation function is observed to scale as $t^{1/2}$, one could undertake a systematic study of surfaces with different average step separations $L$. DSS-limited fluctuations would show a weakly $L$-dependent $G(t)$ while 2D-EC limited fluctuations would be independent of $L$. This is, at the least, a time-consuming experiment and, depending on the specific system, may be unattractive for a host of other technical reasons. An alternative is to evaluate the nature of the correlations of fluctuations between adjacent steps. To accomplish this, another simple statistical analysis has been proposed by Blagojević and Duxbury (BD) to test the physical mechanism governing step fluctuations [23]. Their analysis employs theoretical predictions for the time dependence of the pair correlation function for neighboring steps on a vicinal surface. This function, defined by

$$C_1(t) = \langle x_n(y_0, t+t_0) x_{n+1}(y_0, t_0) \rangle, \quad (5)$$

is also referred to as the cross correlation function because it measures the statistical interdependence of two stochastic processes. If steps on a vicinal surface are approximately isolated, fluctuations of a given step should be nearly independent of the fluctuations of its neighbors in the array. When each step position in Eq. 5 is referenced to *its center of mass* (i.e. $\langle x_n(y,t) \rangle = 0$), $C_1(t)$ should vanish for any isolated step mechanism (i.e. EC, TD, or SED). Additionally, BD show the time dependence that should be observed when step fluctuations are correlated. For neighboring step fluctuations limited by DSS, $C_1(t)$ takes the form[1] [23]:

$$C_1(t) = \sqrt{\frac{16 D c k_B T \Omega^2 t}{9\pi^3 \tilde{\beta} L}} = \frac{1}{6} G_{DSS} \quad . \quad (6)$$

This equation is particularly important because it provides both the form and the magnitude of the correlations that would arise due to DSS. To conclude that DSS is the limiting underlying process, it is necessary to compare measured functions directly with Eq. (6).

As an example of a situation when non-vanishing cross correlations can differ from Eq. 6, we note that while $G(t)$ must *ipso facto* vanish initially, $C_1(t)$ need not. BD's assumption of independently fluctuating steps becomes untenable when the fluctuation amplitude (and thus the wavelength of the measured fluctuations) becomes comparable to $L$; then in-phase fluctuations must dominate, leading to $C_1(t) > 0$. Likewise, there should be in-phase fluctuations when there are strong repulsions between steps (e.g. elastic repulsions and very small step spacings). An upper bound on this effect can be gleaned from the examination in Ref. [13] of the variance of the terrace width distribution in the limit of asymptotically large repulsion, where the tendency to in-phase fluctuations should be strongest. If the steps fluctuate independently, then this variance should be *twice* that deduced in a mean-field, single-active-step model [13], while the calculated value of ~1.8 (rather than 2) indicates a modest reduction due to in-phase step fluctuations. It is unclear precisely how $C_1(t)$ evolves at small times when it is initially finite, but this issue is beyond the scope of this paper, which treats data for which $C_1(t)$ manifestly vanishes initially.

---

[1] In contrast to our conventional definition of $G$ in Eq. (1), BD include a prefactor of ½ on the right-hand side. Hence, the factor of 1/6 in Eq. (6) is 1/3 in their notation.

We have evaluated pair correlations on a surface consisting of a uniform $\sqrt{3} \times \sqrt{3}$ R30° overlayer of Al on Si(111) with an average step separation of 45 nm [20]. Experimental details can be found in Ref. 20. In that work, the temporal correlation function yielded an exponent of $z = 2$, and a number of arguments [20-22] were made to establish the consistency of the EC mechanism. Here we have extended that analysis by computing $C_1(t)$ for STM images that have two neighboring monatomic steps in the field of view. An example of one such image is shown in Fig. 2. As reported previously, this type of pseudo-image is obtained by scanning the STM tip repeatedly over a fixed point on the step edge (for times usually ranging from 20 to 100 seconds) so that the digitized step position results in a time series that lends itself to statistical analysis [4,5,20].

In this method of time-dependent imaging, two adjacent steps are sampled at slightly different times. For example, the step on the right in Fig. 2 only came into view after the step on the left. Thus, the times used in the computation of pair correlations have a small offset. For the image in Fig. 2, the scan rate was 4900 nm/s, and the steps were separated by ~55nm. Thus, the time difference between sampling adjacent steps was only ~11 ms, which is a factor of 4 smaller than the temporal resolution along the vertical axis of the pseudo-image (i.e. the temporal resolution with which G(t) can be measured). As long as the offset is constant for all images (e.g. the variability in step spacing is much smaller than the net step spacing), this does not create a problem for the analysis.

The results for *G(t)* and *C₁(t)* are shown in Fig. 3a and 3b, respectively, for four images obtained at a temperature of 970 K (note the expanded scale in Fig. 3b). The curves in Fig. 3 were obtained by averaging curves from 4 STM images, each with two adjacent steps. The total imaging time in each case was 23 seconds, but the curves are only plotted for the first few seconds when *G(t)* shows unmistakable power-law growth and comparison with the predictions of BD is straightforward. However, for the full 23 second measurement time the pair correlation function fluctuates around zero as well .

There is no obvious functional form to the time dependence of the measured pair correlation function. The statistical error in the curve in Fig. 3b is consistent with vanishing pair correlations. For a curve that is identically zero, according to the predictions of Ref. 23, there are no correlated step dynamics, and the steps can be considered as effectively isolated in their fluctuations.

Direct comparison with Eq. (6) further establishes the dominant isolated step behavior for this surface. Fig. 4 shows 1/6 of the measured correlation function from Fig. 3a on the same axes as the corresponding measured pair correlation function from Fig. 3b. The experiment does not match the DSS prediction that $C_1(t) = \frac{1}{6} G(t)$ well, although for times less than about 1.5 sec the large statistical scatter in the individual points in Fig. 3b does intersect the 1/6 G curve in Fig. 4. This large scatter in the cross correlation function is simply the result of the fact that, for any given observation, the steps are equally likely to be moving either in phase or out of phase when the measurement begins. To make a better quantitative comparison, the experimental curves were each fit to a two-parameter power law and the results plotted as solid lines in Fig. 4. The fit results were

$$\frac{1}{6}G(t) = (33 \pm 1) \cdot t^{(0.41 \pm 0.03)} \text{ and } C_1(t) = (-32 \pm 6) \cdot t^{(0.26 \pm 0.17)}.$$ The parameters for $G(t)$ are in agreement with the results of Ref. 20 (not surprising, since the data presented here were included in that work) while the parameters for $C_1(t)$ are statistically inconsistent with Eq. 6. Thus, the approximate $t^{\frac{1}{2}}$ scaling of $G(t)$ cannot be attributed to DSS kinetics.

In summary, we have extracted the nearest-neighbor pair correlation function for steps on Si(111)-$\sqrt{3} \times \sqrt{3}$ R30°-Al and found that it is not significantly different from zero. Thus, the stochastic evolution of neighboring steps is nearly statistically independent [17], eliminating DSS as a relaxation mode and providing strong positive evidence for evaporation-condensation at the step edges as the rate-limiting relaxation mode governing fluctuations. This represents a more controlled experimental procedure for evaluating the DSS mechanism than testing the step separation-dependence of Eq. (4) using different samples prepared with different step densities. In this way, from a single experiment on a vicinal surface, it is possible to extract enough information to distinguish between the DSS and 2D-EC mechanisms of step relaxation.

The evidence presented here strongly supports the interpretation of EC-limited step fluctuations on the Si(111) surface with a $\sqrt{3} \times \sqrt{3}$ R30° overlayer of Al.

Interpreting fluctuation kinetics in this way has met with some criticism in the literature based on simple models for the activated process of step attachment [13,14]. Specifically, using a rigid-lattice substrate, modeling attachment with a one-dimensional activation barrier, and requiring the same pre-exponential factor as for terrace diffusion leads to a larger activation barrier for EC compared with the activation barrier for terrace diffusion. However, such simple models do not account for cooperative rebonding at the step edge [24, 25] during the attachment process and its possible consequences for step edge kinetics [26]. They also do not account for the changes in the pre-exponential factor that will arise when transport is influenced by step vibrational modes [27] or complicated exchange mechanisms [28, 29]. In the case of the $\sqrt{3} \times \sqrt{3}$ R30° overlayer of Al on Si(111), both rebonding and complicated transition states are expected. The simple reasons generally given for rejecting step attachment (2D-EC) as rate-limiting can reasonably be set aside for this system in the face of the strong supporting evidence presented here and elsewhere [20]. From a more general perspective, the analysis described here (originally developed in Ref 23) may help to clarify mass transport in the many other systems where ambiguity exists regarding the role of DSS or 2D-EC processes. We expect that in such further analyses the issues of crossover between competing atomistic processes and the range of fluctuation wavelengths sampled (i.e. effective system size) will be crucial to a correct interpretation of the observations [30].

The analysis presented in the present work provides a useful tool for addressing continuum step issues. To go beyond this phenomenological understanding, atomistic models of the kinetic processes of step attachment for complicated surface reconstructions would be of great value.

This work was supported by UMD-NSF-MRSEC under DMR-00-80008.

Figure Captions

Fig. 1   Schematic illustration of the two surface processes, evaporation-condensation at an isolated step, and diffusion across the terraces between steps, leading to $t^{1/2}$ scaling of the temporal correlation function.

Fig. 2   STM "pseudo-image" obtained at 970 K by scanning the tip repeatedly over a fixed position on the step edge ($y$). The horizontal direction in the image is the x-direction and the vertical axis corresponds to the total duration of the scan. Thus, the dimensions of the image are (100 nm × 23 s).

Fig. 3   a.) The usual temporal correlation function averaged over 4 images like the one shown in Fig. 2 at 970 K. The computed curves are only shown for early times when power law scaling is expected. b.) The pair correlation function averaged over the same four images used to measure the curve in part a.) and plotted over the identical time span. Note the change in vertical scale between a.) and b.). Error bars are the standard error from the four separate measurements.

Fig. 4   Comparison of the average of the pair correlation functions (open circles) with the theoretical prediction of 1/6 of the value of the temporal correlation function G(t) (closed circles) to test the prediction for DSS from Ref. 17. The thick solid curves are the best fits to a two-parameter power law of the form shown in Eq. (2) of the text. The pair correlation function is not well-fit by a power law. Furthermore the prediction that the two quantities in this plot should be equal for DSS-limited step kinetics is clearly not satisfied. (Error bars are the same as in Fig. 3 but omitted for clarity of the fit lines).

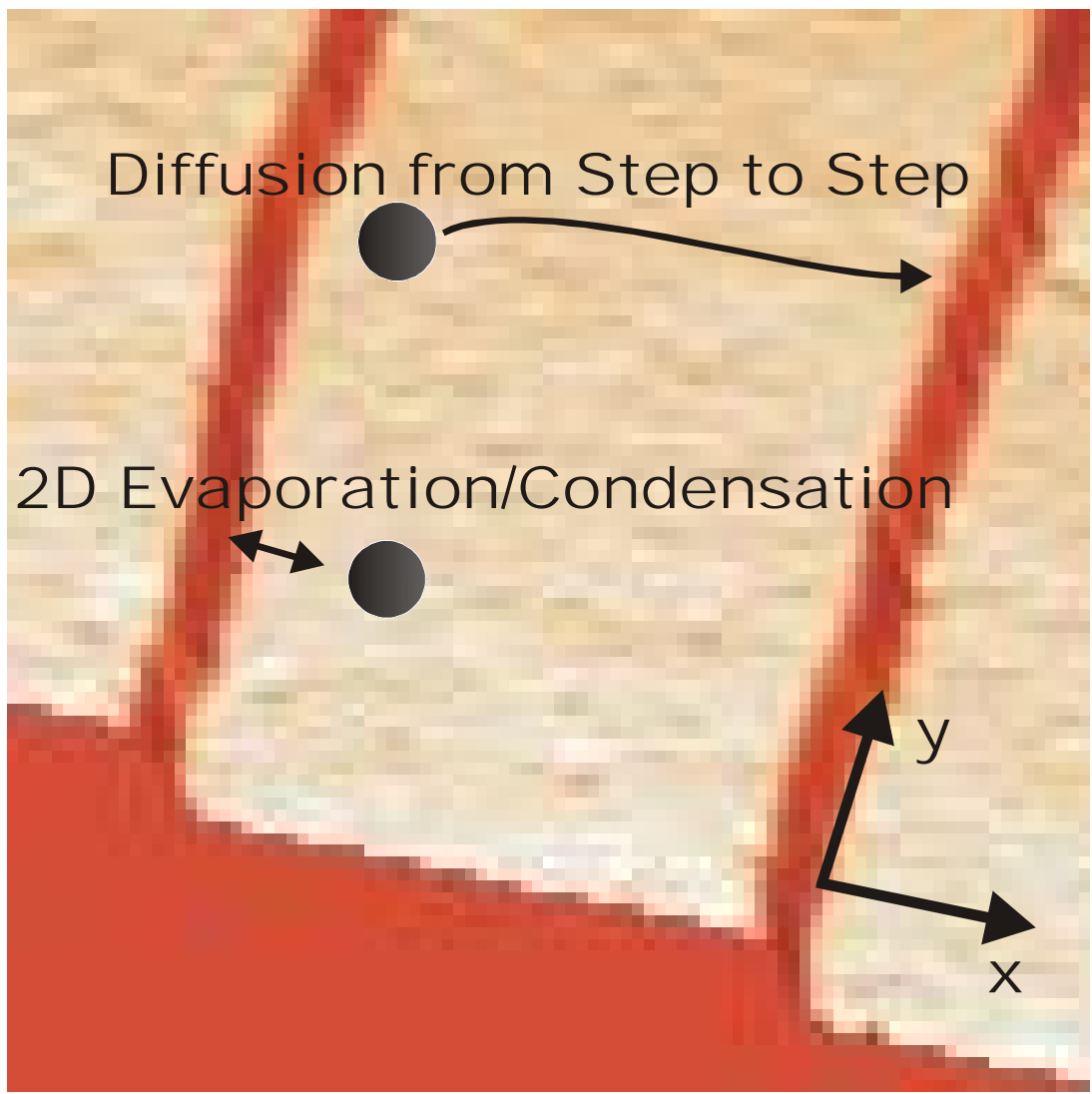

Fig. 1

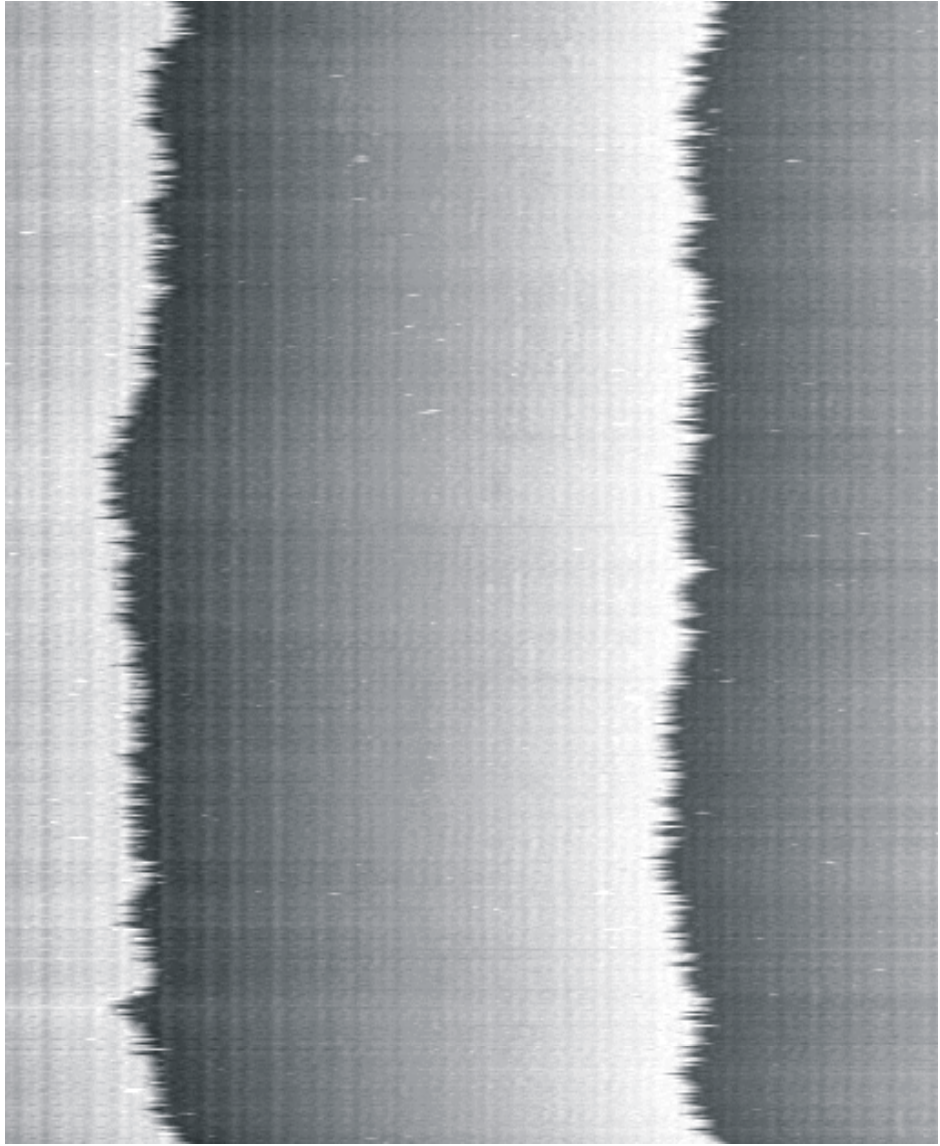

Fig. 2

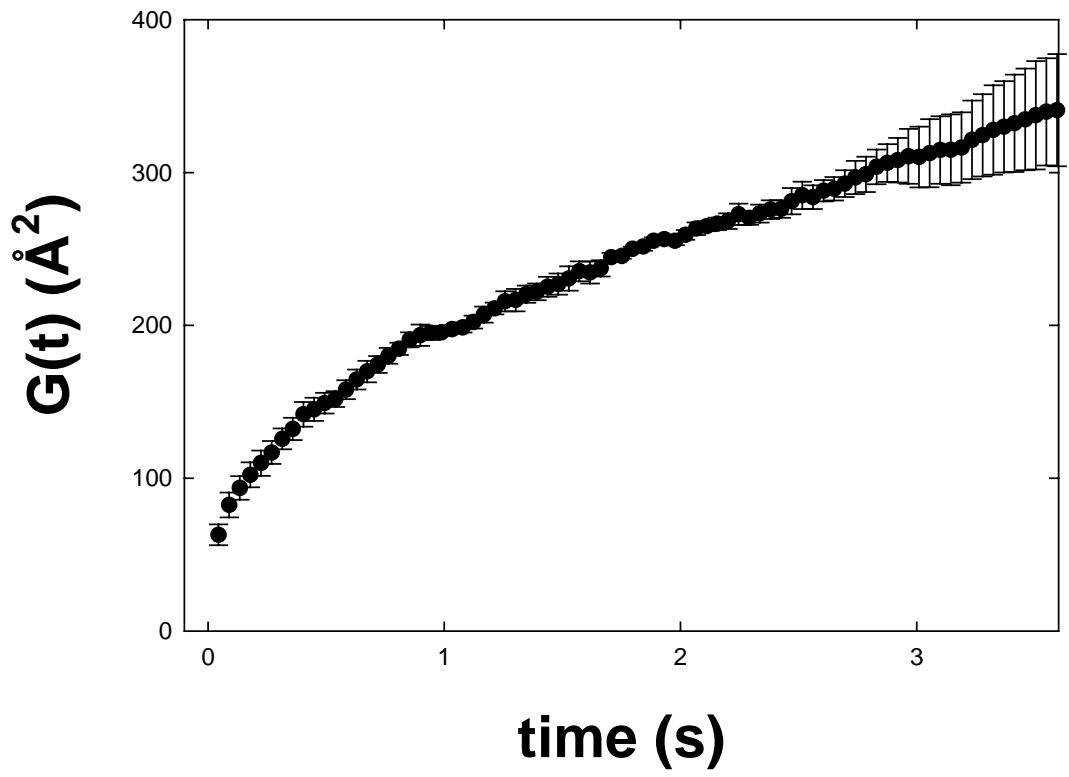
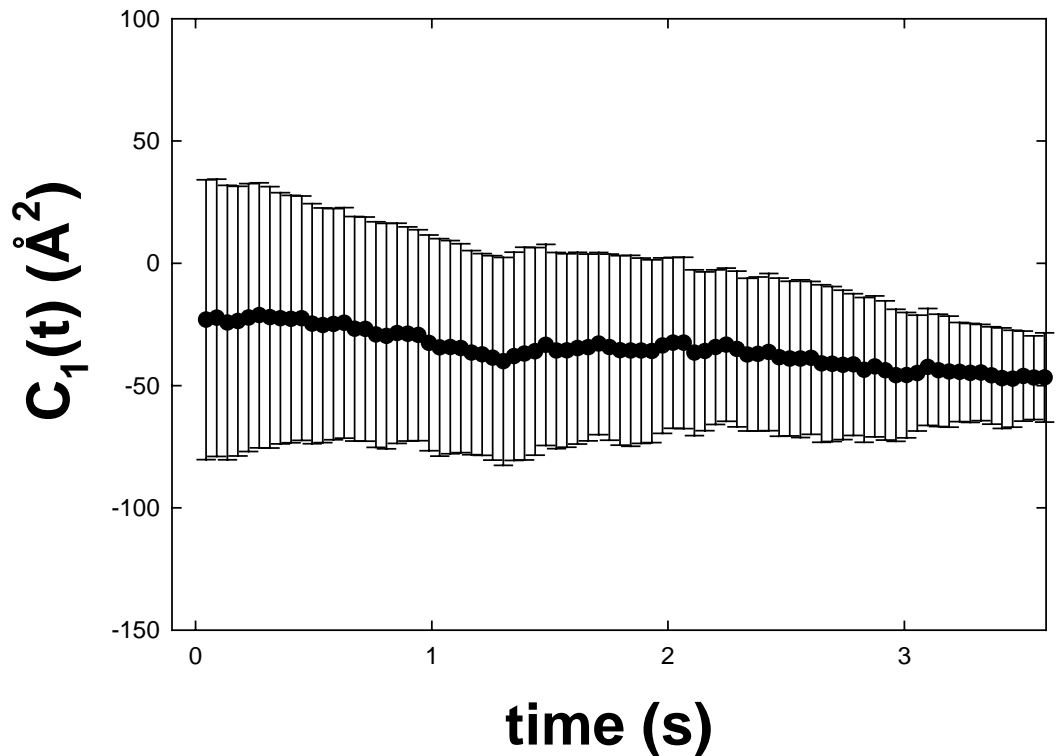

Fig.3

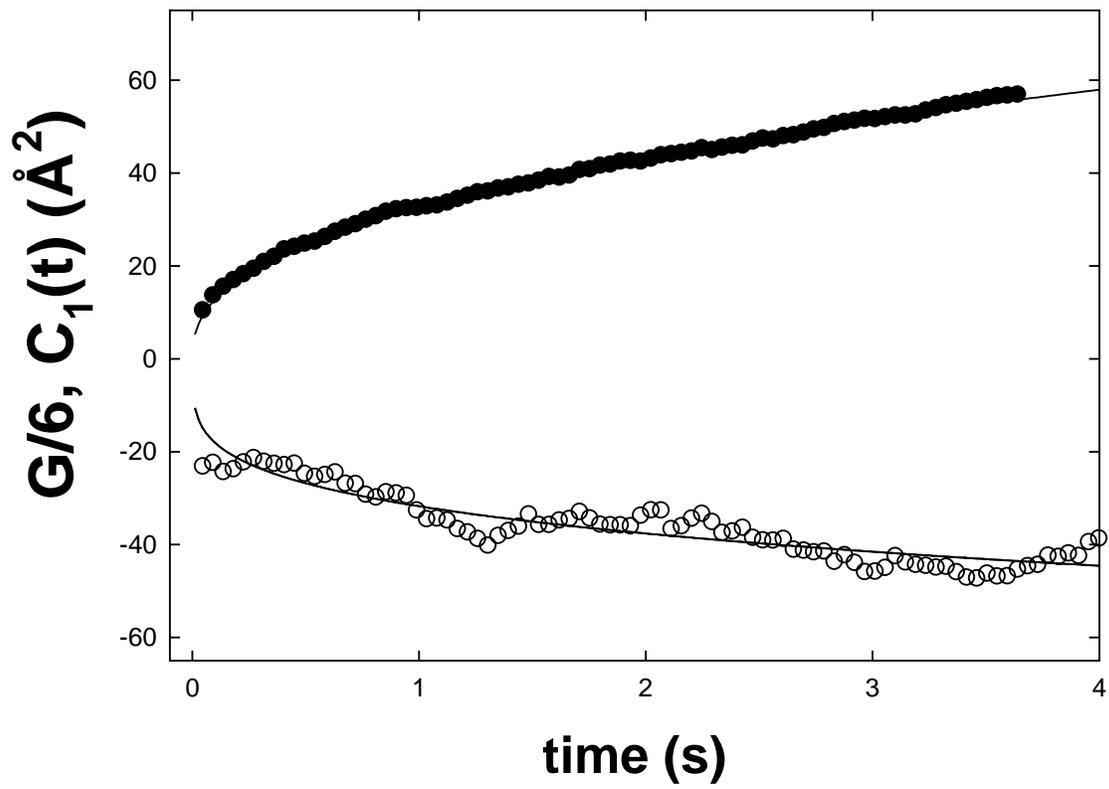

Fig. 4